\documentclass[11pt]{article}
\usepackage{amsmath,amssymb}
\usepackage{graphicx}
\usepackage{subcaption}
\usepackage{xcolor}
\usepackage{caption}

\begin{document}

\setcounter{page}{1}

\pagestyle{plain}

\begin{center}
\Large{\bf Primordial black holes and induced gravitational waves from localized features in DBI inflation}\\
\small \vspace{1cm} {Narges
Rashidi}\footnote{n.rashidi@umz.ac.ir } \\
\vspace{0.5cm} $^{}$ Department of Theoretical Physics, Faculty of
Science,
University of Mazandaran,\\
P. O. Box 47416-95447, Babolsar, IRAN\\
\end{center}

\begin{abstract}
We investigate primordial black hole formation in a Dirac-Born-Infeld inflationary framework in which the background dynamics are determined by explicit functional forms of the inflaton potential $V(\phi)$ and the warp factor $f(\phi)$. The background equations are solved numerically to obtain the evolution of the slow-roll parameters. We show that a localized feature in the potential, accompanied by a correlated structure in the warp factor, dynamically induces a transient suppression of the slow-roll parameter, leading to a short-lived non-attractor phase. This behavior generates an enhancement of the curvature power spectrum on small scales, while preserving consistency with CMB-scale observables. The Mukhanov-Sasaki equation is then solved numerically to compute the resulting power spectrum, which exhibits narrow and localized peaks in the PBH abundance across different mass ranges. We also evaluate the associated stochastic background of induced gravitational waves and find that the predicted signal can fall within the sensitivity bands of future pulsar timing arrays and space-based interferometers, depending on the scale of the inflationary feature. A parameter scan in the $(r,n_s)$ plane shows that the large-scale predictions remain consistent with current observational constraints.
\end{abstract}

\newpage

\section{Introduction}

The physics of the inflationary epoch on small scales remains largely unconstrained and poorly understood. While observations of the cosmic microwave background (CMB) provide precise information about primordial curvature perturbations on large cosmological scales, revealing an almost scale-invariant, Gaussian, and adiabatic spectrum~\cite{pl18b} - they probe only a limited window of the inflationary dynamics. On much smaller scales, which are not directly accessible to CMB anisotropies or conventional large-scale structure measurements, the primordial power spectrum is far less constrained, leaving room for departures from an approximately scale-invariant form. In particular, a sufficiently strong enhancement of curvature perturbations can seed large overdensities that gravitationally collapse after horizon re-entry, leading to the formation of primordial black holes (PBHs)~\cite{Car74,Car75,Sa18}. Owing to their purely gravitational interactions and their broad mass spectrum, PBHs constitute a compelling non-particle candidate for dark matter and may also contribute to the population of black holes observed through gravitational wave experiments\cite{Ze67,Kh10,Ka19,Gr21,Car20}. The production of PBHs in inflationary scenarios typically requires a temporary departure from standard slow-roll evolution, motivating the exploration of inflationary models capable of naturally generating localized features in the small-scale power spectrum.

Within the concordance $\Lambda$CDM framework, a substantial fraction of the present-day energy density is in the form of cold dark matter, whose microscopic nature remains unknown despite increasingly precise cosmological measurements~\cite{pl18cp}. PBHs offer a qualitatively different possibility: rather than introducing new particle degrees of freedom, they arise from gravitational collapse in the early Universe and can behave as effectively collisionless matter over cosmological timescales. A wide variety of astrophysical and cosmological probes - including CMB bounds, lensing searches, dynamical effects, accretion signatures, and gravitational wave observations - place strong limits on the PBH abundance across many mass scales, while leaving particular regions of parameter space still viable~\cite{Gr21,Car20}. This motivates inflationary scenarios in which the small-scale curvature perturbations are amplified in a controlled and localized way, potentially yielding an appreciable PBH population consistent with current constraints~\cite{Oz23}.

Realizing a sizable PBH abundance within inflationary cosmology requires a transient amplification of primordial curvature perturbations on scales far smaller than those directly constrained by CMB and large-scale structure observations. Such enhancements typically arise from a temporary violation of the standard slow-roll conditions and can be achieved through a variety of mechanisms, including the presence of inflection points or localized features in the inflationary potential, transient ultra-slow-roll phases, or additional dynamical degrees of freedom. In particular, inflationary models with non-canonical kinetic terms offer a compelling framework, as variations in the effective sound speed can efficiently amplify curvature perturbations while maintaining consistency with large-scale observational constraints~\cite{Ka19,Oz23}. 

Among the various inflationary mechanisms capable of enhancing primordial fluctuations on small scales, scenarios with non-canonical kinetic structures offer a particularly flexible and physically motivated framework for PBH formation. In the DBI framework, this behavior is determined by the inflaton potential and the warp factor, which together control the background dynamics. As a result, localized features in these functions can naturally induce short-lived departures from slow-roll evolution and lead to controlled enhancements of the curvature power spectrum.

In this work, we investigate a Dirac–Born–Infeld (DBI) inflationary model by specifying explicit functional forms for the inflaton potential \(V(\phi)\) and the warp factor \(f(\phi)\). The background equations are solved numerically, and the slow-roll parameters, scalar spectral index, tensor-to-scalar ratio, the primordial curvature power spectrum, the PBH abundance, and the induced gravitational waves (IGWs) spectrum are obtained as outputs of the model. We show that a localized feature in the potential, together with a correlated structure in the warp factor, dynamically generates a transient suppression of the slow-roll parameter. This establishes a direct connection between the DBI functions \(V(\phi)\), \(f(\phi)\), and the small-scale amplification of curvature perturbations, while preserving consistency with CMB-scale observables. The rest of this paper is organized as follows. In Sec.~\ref{sec2}, we introduce the DBI action, specify the potential and warp factor, and derive the background evolution equations. In Sec.~\ref{sec3p1}, we present the scalar perturbation dynamics in the presence of a time-dependent sound speed and compute the primordial curvature power spectrum. In Sec.~\ref{sec3p2}, we connect the resulting enhancement of \(\mathcal{P}_\zeta(k)\) to PBH formation and summarize the expressions for the PBH abundance. In Sec.~\ref{sec4}, we present the numerical results, including the background evolution, the emergent slow-roll and sound-speed profiles, the enhanced power spectrum, the PBH abundance, and the associated IGWs signal. We also confront the large-scale predictions with current observational constraints in the \(r-n_s\) plane. Finally, we conclude in Sec.~\ref{sec5}.

\section{DBI framework and background evolution}
\label{sec2}

The DBI inflationary scenario arises from string-theoretic constructions, where the inflaton field can be interpreted as the position of a D-brane moving in a warped extra-dimensional background. In this framework, the non-canonical kinetic structure modifies the background evolution of the inflaton field, leading to a nontrivial behavior of the slow-roll parameters. In particular, the dynamics of the first slow-roll parameter are directly influenced by the form of the warp factor $f(\phi)$ and the inflaton potential $V(\phi)$. This feature plays an important role in generating departures from standard slow-roll evolution and can significantly affect the amplification of primordial curvature perturbations. In this regard, we consider the following effective DBI-type action
\begin{equation}
	S = \int d^4x \sqrt{-g}
	\left[
	\frac{M_{\rm Pl}^2}{2} R
	-\frac{1}{f(\phi)}
	\left(\sqrt{1-2f(\phi)X}-1\right)
	- V(\phi)
	\right],
	\qquad
	X \equiv -\frac{1}{2} g^{\mu\nu}\partial_\mu\phi\partial_\nu\phi .
	\label{eq1}
\end{equation}
Here \(M_{\rm Pl}\) denotes the reduced Planck mass, \(R\) is the Ricci scalar constructed from the spacetime metric \(g_{\mu\nu}\), and \(\phi\) is the inflaton field. The function \(f(\phi)\) acts as an effective warp factor controlling the non-canonical kinetic structure, while \(V(\phi)\) denotes the inflaton potential. In this work, the DBI action is treated as a phenomenological effective description, without assuming a specific ultraviolet completion. We use reduced Planck units with \(M_{\rm Pl}=1\), while dimensional quantities are quoted explicitly in units of the reduced Planck mass where appropriate.

The non-canonical kinetic term implies a reduced propagation speed for scalar perturbations,
\begin{equation}
	c_s^2 = 1 - 2 f(\phi) X ,
	\label{eq2}
\end{equation}
and it is convenient to introduce the associated Lorentz factor
\begin{equation}
	\gamma \equiv \frac{1}{c_s}.
	\label{eq3}
\end{equation}

We assume a spatially flat Friedmann-Lema\^{i}tre-Robertson-Walker (FLRW) background with line element
\begin{equation}
	ds^2 = -dt^2 + a^2(t)\,\delta_{ij}dx^i dx^j,
	\label{eq4}
\end{equation}
where $a(t)$ is the scale factor and $H \equiv \dot a/a$ denotes the Hubble expansion rate. We define the number of e-folds as $N \equiv \ln a$.

The background dynamics are governed by the Friedmann equation
\begin{equation}
	3 M_{\rm Pl}^2 H^2 = \rho ,
	\label{eq5}
\end{equation}
where the energy density of the DBI field is given by
\begin{equation}
	\rho = \frac{\gamma - 1}{f(\phi)} + V(\phi).
	\label{eq6}
\end{equation}

The evolution of the inflaton field is obtained by solving the corresponding equation of motion derived from the action,
\begin{equation}
	\ddot{\phi} + 3H \frac{1}{\gamma^2}\dot{\phi} + \frac{V_{,\phi}}{\gamma^3}
	+ \frac{f_{,\phi}}{2f^2}\left(1 - \frac{1}{\gamma^3}\right) = 0,
	\label{eq7}
\end{equation}
together with the Friedmann equation~\eqref{eq5}.

In this work, we specify explicit functional forms for the inflaton potential $V(\phi)$ and the warp factor $f(\phi)$, given by
\begin{equation}
	V(\phi) = V_0 \left(1 - e^{-\beta \phi} \right)^2
	\left[
	1 - A_{\rm V}\,
	\exp\!\left(-\frac{(\phi-\phi_{0})^2}{2\sigma^2}\right)
	\right],
	\label{eq8}
\end{equation}
and
\begin{equation}
	f(\phi) = f_0\, 
	\left[
	1 + A_{f}\,
	\exp\!\left(-\frac{(\phi-\phi_{0})^2}{2\sigma^2}\right)
	\right].
	\label{eq9}
\end{equation}
Here \(\phi_0\) and \(\sigma\) are measured in units of \(M_{\rm Pl}\), \(\beta\) in units of \(M_{\rm Pl}^{-1}\), and \(f_0\) in units of \(M_{\rm Pl}^{-4}\).

The potential in Eq.~\eqref{eq8} consists of a plateau-like term of the form $(1-e^{-\beta\phi})^2$, which supports slow-roll inflation at large field values, modulated by a localized Gaussian dip. The parameter $V_0$ sets the overall energy scale of inflation, while $\beta$ controls the slope of the plateau. The amplitude $A_{\rm V}$ determines the depth of the localized feature, $\phi_{0}$ specifies its position in field space, and $\sigma$ controls its width. This localized structure induces a temporary flattening of the potential, leading to a transient suppression of the slow-roll parameter $\epsilon_1$. Similarly, the warp factor in Eq.~\eqref{eq9} consists of a constant background value $f_0$ modulated by a localized Gaussian enhancement. The parameter $A_f$ controls the strength of this enhancement, while its position and width are determined by the same parameters $\phi_{0}$ and $\sigma$. This choice ensures that the variation in the warp factor is correlated with the feature in the potential. 
The combined effect of the dip in the potential and the enhancement in the warp factor dynamically generates a short-lived departure from the slow-roll regime, resulting in a temporary amplification of curvature perturbations on the corresponding scales.

We introduce the first two slow-roll parameters as
\begin{equation}
	\epsilon_1 \equiv -\frac{\dot H}{H^2}
	= -\frac{1}{H}\frac{dH}{dN},
	\label{eq10}
\end{equation}
and
\begin{equation}
	\epsilon_2 \equiv \frac{d\ln \epsilon_1}{dN}.
	\label{eq11}
\end{equation}

In addition to the slow-roll parameters, the variation of the sound speed is characterized by
\begin{equation}
	s \equiv \frac{d\ln c_s}{dN}.
	\label{eq12}
\end{equation}

Localized features in the potential $V(\phi)$ and the warp factor $f(\phi)$ can dynamically induce a transient suppression of $\epsilon_1$. Such behavior corresponds to a short-lived departure from standard slow-roll evolution and plays a central role in enhancing the curvature power spectrum on small scales.

In models with a varying sound speed, the scalar spectral index is generally given by
\begin{equation}
	n_s - 1 = -2\epsilon_1 - \epsilon_2 - s ,
	\label{eq13}
\end{equation}
evaluated at sound-horizon crossing. 
The tensor-to-scalar ratio is correspondingly modified and takes the form
\begin{equation}
	r = 16\,\epsilon_1\,c_s ,
	\label{eq14}
\end{equation}
also evaluated at sound-horizon crossing.

The background quantities obtained in this section provide the necessary input for the evolution of scalar perturbations. In the following section, we analyze the curvature perturbations by solving the Mukhanov–Sasaki equation and study the resulting enhancement of the primordial power spectrum on small scales.

\section{Scalar perturbations and primordial power spectrum}

\subsection{Scalar perturbations}
\label{sec3p1}

We now study scalar perturbations around the DBI background constructed in Sec.~\ref{sec2}. 
For a single-field model with a time-dependent sound speed, the dynamics of the comoving curvature perturbation (here denoted by $\zeta$) are captured by the quadratic action~\cite{Muk88,Sas86}
\begin{equation}
	S^{(2)}=\frac{1}{2}\int d\tau\, d^3x\; z^2
	\left[
	(\zeta')^2-c_s^2(\nabla\zeta)^2
	\right],
	\label{eq15}
\end{equation}
where a prime denotes differentiation with respect to conformal time $\tau$ defined via $d\tau=dt/a(t)$. 
The background-dependent normalization is
\begin{equation}
	z \equiv \frac{a\sqrt{2\epsilon_1}\,M_{\rm Pl}}{c_s}.
	\label{eq16}
\end{equation}

Introducing the canonically normalized variable
\begin{equation}
	u \equiv z\,\zeta,
	\label{eq17}
\end{equation}
the Fourier modes $u_k$ satisfy the Mukhanov-Sasaki equation
\begin{equation}
	u_k''+\left(c_s^2 k^2-\frac{z''}{z}\right)u_k=0.
	\label{eq18}
\end{equation}
In the DBI setup considered here, the slow-roll parameters are obtained from the numerical solution of the background equations, determined by the specified potential and warp factor. The resulting profiles can exhibit localized features, which induce nontrivial structure in the effective term $z''/z$, making numerical evolution of Eq.~\eqref{eq18} necessary in the feature regime.

We impose the Bunch--Davies vacuum deep inside the sound horizon,
\begin{equation}
	u_k(\tau)\;\longrightarrow\;\frac{1}{\sqrt{2c_s k}}
	\exp\!\left(-i c_s k\tau\right),
	\qquad c_s k \gg aH,
	\label{eq19}
\end{equation}
and evolve until the mode freezes on super-horizon scales. The primordial curvature power spectrum is then computed as
\begin{equation}
	\mathcal{P}_{\zeta}(k)=\frac{k^3}{2\pi^2}\,|\zeta_k|^2
	=\frac{k^3}{2\pi^2}\left|\frac{u_k}{z}\right|^2,
	\label{eq20}
\end{equation}
evaluated at late times after horizon exit.

\subsection{Power spectrum enhancement and PBH formation}
\label{sec3p2}

The DBI background introduced in Sec.~\ref{sec2}, together with the scalar perturbation dynamics discussed above, can allow for a transient enhancement of curvature perturbations on scales much smaller than those probed by the CMB. Such an enhancement is induced by localized features in the inflaton potential $V(\phi)$ and the warp factor $f(\phi)$, which lead to a transient suppression of the slow-roll parameters. These effects modify the effective mass term $z''/z$ in the Mukhanov--Sasaki equation~\eqref{eq18}. As a result, curvature modes can experience a period of growth before freezing on super-horizon scales, leading to a pronounced peak in the primordial power spectrum.

Once the enhanced power spectrum $\mathcal{P}_{\zeta}(k)$ is obtained by numerically solving Eq.~\eqref{eq18}, it can be used to estimate the abundance of PBHs formed after horizon re-entry during the radiation-dominated era. The key quantity controlling PBH formation is the variance of the coarse-grained density contrast, which can be expressed as~\cite{You14}
\begin{equation}
	\sigma^2(k_0)
	=
	\int_0^\infty \frac{dk}{k}\,
	W^2\!\left(\frac{k}{k_0}\right)\,
	\mathcal{P}_{\delta}(k),
	\label{eq21}
\end{equation}
where $k_0$ is the comoving wavenumber defining the smoothing scale for the coarse-grained density contrast (typically chosen around the scale of interest), and $W(k/k_0)$ is a window function smoothing perturbations on the comoving scale $k_0^{-1}$. In the radiation-dominated epoch, the density contrast at horizon re-entry can be related to the curvature perturbation through
\begin{equation}
	\mathcal{P}_{\delta}(k)
	=
	\frac{16}{81}
	\left(\frac{k}{k_0}\right)^4
	\mathcal{P}_{\zeta}(k),
	\label{eq22}
\end{equation}
so that the variance becomes directly sensitive to the small-scale enhancement of $\mathcal{P}_{\zeta}(k)$.

Assuming a Gaussian distribution for the primordial fluctuations, the fraction of the Universe collapsing into PBHs at formation is given by~\cite{Sa18,Shi99,Pol07,Mus09,Har13}
\begin{equation}
	\beta (M)
	=
	\gamma_c
	\int_{\delta_c}^{\infty}
	\frac{d\delta}{\sqrt{2\pi}\sigma(M)}
	\exp\!\left(-\frac{\delta^2}{2\sigma^2(M)}\right)
	\simeq
	\frac{\gamma_c\,\sigma(M)}{\sqrt{2\pi}\,\delta_c}
	\exp\!\left(-\frac{\delta_c^2}{2\sigma^2(M)}\right),
	\label{eq23}
\end{equation}

where $\delta_c$ denotes the threshold for gravitational collapse and $\gamma_c$ parametrizes the efficiency of PBH formation. Following standard estimates in the radiation-dominated era, we adopt $\delta_c\simeq0.4$ and $\gamma_c\simeq0.2$.

The PBH mass associated with a comoving scale $k$ is estimated by the horizon mass at re-entry,
\begin{equation}
	M(k)
	\simeq
1.13\times 10^{15}	\gamma_c\, M_\odot
	\left(\frac{g_*}{106.75}\right)^{-1/6}
	\left(\frac{k}{k_*}\right)^{-2},
	\label{eq24}
\end{equation}
where $g_*$ is the effective number of relativistic degrees of freedom at formation and $k_*=0.05\,{\rm Mpc}^{-1}$ denotes the CMB pivot scale. We use $g_*=106.75$, corresponding to the Standard Model value at high temperatures.

The present-day fractional abundance of PBHs is then given by~\cite{Sa18}
\begin{equation}
	f_{\rm PBH}(M)
	\simeq
	\frac{\Omega_{\rm PBH}(M)}{\Omega_{\rm DM}}
	\simeq
	1.6\times10^8
	\left(\frac{\gamma_c}{0.2}\right)^{1/2}
	\left(\frac{g_*}{106.75}\right)^{-1/4}
	\left(\frac{M}{M_\odot}\right)^{-1/2}
	\beta (M),
	\label{eq25}
\end{equation}
where the proportionality factor accounts for the redshifting of PBHs relative to radiation.

In the DBI scenario considered here, the localized enhancement of $\mathcal{P}_{\zeta}(k)$ generated during inflation directly translates into a peaked PBH abundance. This allows for a controlled exploration of PBH abundances across a wide mass range while maintaining consistency with CMB-scale constraints, as demonstrated in the numerical analysis presented in the next section.

\subsection{Induced gravitational waves}
\label{sec3gw}

An enhancement of the scalar curvature perturbations on small scales not only leads to the formation of PBHs, but also inevitably generates a stochastic background of gravitational waves at second order in cosmological perturbation theory. These tensor modes, commonly referred to as IGWs, provide an important complementary probe of small-scale inflationary physics. At second order, the tensor perturbations are sourced by quadratic combinations of scalar fluctuations. As a result, the evolution equation for the tensor modes contains a source term constructed from the scalar perturbations. Solving this equation during the radiation-dominated era, the gravitational wave energy density at horizon re-entry can be expressed as a convolution integral over the scalar power spectrum.
The present-day gravitational wave energy density spectrum is defined as
$\Omega_{\rm GW,0}(f) \equiv \frac{1}{\rho_c}\frac{d\rho_{\rm GW}}{d\ln f}
$
where $\rho_c$ is the critical energy density of the Universe. This quantity provides a convenient dimensionless measure of the gravitational wave background and allows for a direct comparison with observational sensitivities.
More specifically, the dimensionless energy density spectrum at the time of horizon crossing, $\Omega_{\rm GW}(\eta_c,k)$, can be written as (see e.g.~\cite{Ananda07,Baumann07,Kohri18})
\begin{equation}
	\Omega_{\rm GW}(\eta_c,k)
	=
	\frac{1}{12}
	\int_0^\infty dv
	\int_{|1-v|}^{1+v} du\,
	\left[
	\frac{4v^2-(1+v^2-u^2)^2}{4uv}
	\right]^2
	\mathcal{P}_{\zeta}(ku)\,
	\mathcal{P}_{\zeta}(kv)\,
	I^2(u,v),
	\label{eq26}
\end{equation}
where $u$ and $v$ are dimensionless integration variables. The first bracket encodes the geometric coupling between different modes, while the function $I(u,v)$ accounts for the time evolution of the source term during the radiation era. This function can be written as
\begin{eqnarray}
	I^2(u,v)
	=
	\left[
	\frac{3(u^2+v^2-3)}{4u^3 v^3}
	\left(
	-4uv + (u^2+v^2-3)\ln\left|\frac{3-(u+v)^2}{3-(u-v)^2}\right|
	\right)
	\right]^2\nonumber\\
	+
	\pi^2
	\left[
	\frac{3(u^2+v^2-3)}{4u^3 v^3}
	\right]^2
{\cal{H}}(u+v-\sqrt{3}),
	\label{eq27}
\end{eqnarray}
where ${\cal{H}}$ denotes the Heaviside step function, reflecting the resonance condition in the source term.

To relate the gravitational wave spectrum at horizon crossing to the present time, one must account for the redshift and the change in the number of relativistic degrees of freedom. The present-day energy density spectrum is given by
\begin{equation}
	\Omega_{\rm GW,0}(k) h^2
	=
	0.83
	\left(\frac{g_c}{10.75}\right)^{-1/3}
	\Omega_{r,0} h^2\,
	\Omega_{\rm GW}(\eta_c,k),
	\label{eq28}
\end{equation}
where $\Omega_{r,0} h^2 \simeq 4.2\times10^{-5}$ is the current radiation density and $g_c$ denotes the effective number of relativistic degrees of freedom at the time of production. In the present DBI setup, the localized enhancement of the scalar power spectrum leads to a peaked gravitational wave signal, whose amplitude and frequency are directly determined by the position and shape of the feature in $\mathcal{P}_{\zeta}(k)$.

\section{Numerical results and observational constraints}
\label{sec4}
\subsection{Background evolution and slow-roll dynamics}
In this section, we present the numerical implementation of the DBI model developed in the previous sections. The present analysis starts directly from the potential $V(\phi)$ and the warp factor $f(\phi)$ given in Eqs.~\eqref{eq8} and~\eqref{eq9}. The background equations are solved numerically, and the resulting functions $\epsilon_1(N)$ and $\epsilon_2(N)$ are obtained as derived quantities. These background solutions are then used to solve the Mukhanov--Sasaki equation and to compute the primordial curvature power spectrum, the PBH abundance, and the IGWs spectrum. We also confront the large-scale predictions with current observational constraints on the scalar spectral index and the tensor-to-scalar ratio. The end of inflation is defined by the condition $\epsilon_1=1$, which determines the total number of e-folds. Following standard CMB conventions, we take the pivot scale to be $k_*=0.05\,{\rm Mpc}^{-1}$ and associate it with the mode that exits the sound horizon about $55$ e-folds before the end of inflation. For each parameter set, the overall normalization of the potential is fixed by matching the scalar power spectrum at the pivot scale to the observed amplitude $A_s\simeq 2.1\times10^{-9}$. This procedure determines the absolute energy scale of inflation, while the relative enhancement of the power spectrum on small scales is controlled by the localized structures in $V(\phi)$ and $f(\phi)$. The dimensional parameters appearing in Eqs.~\eqref{eq8} and~\eqref{eq9} are quoted in reduced Planck units, as specified in Table~\ref{tab1}.

\begin{figure}[t]
	\centering
	\begin{subfigure}{0.40\linewidth}
		\centering
		\includegraphics[width=\linewidth]{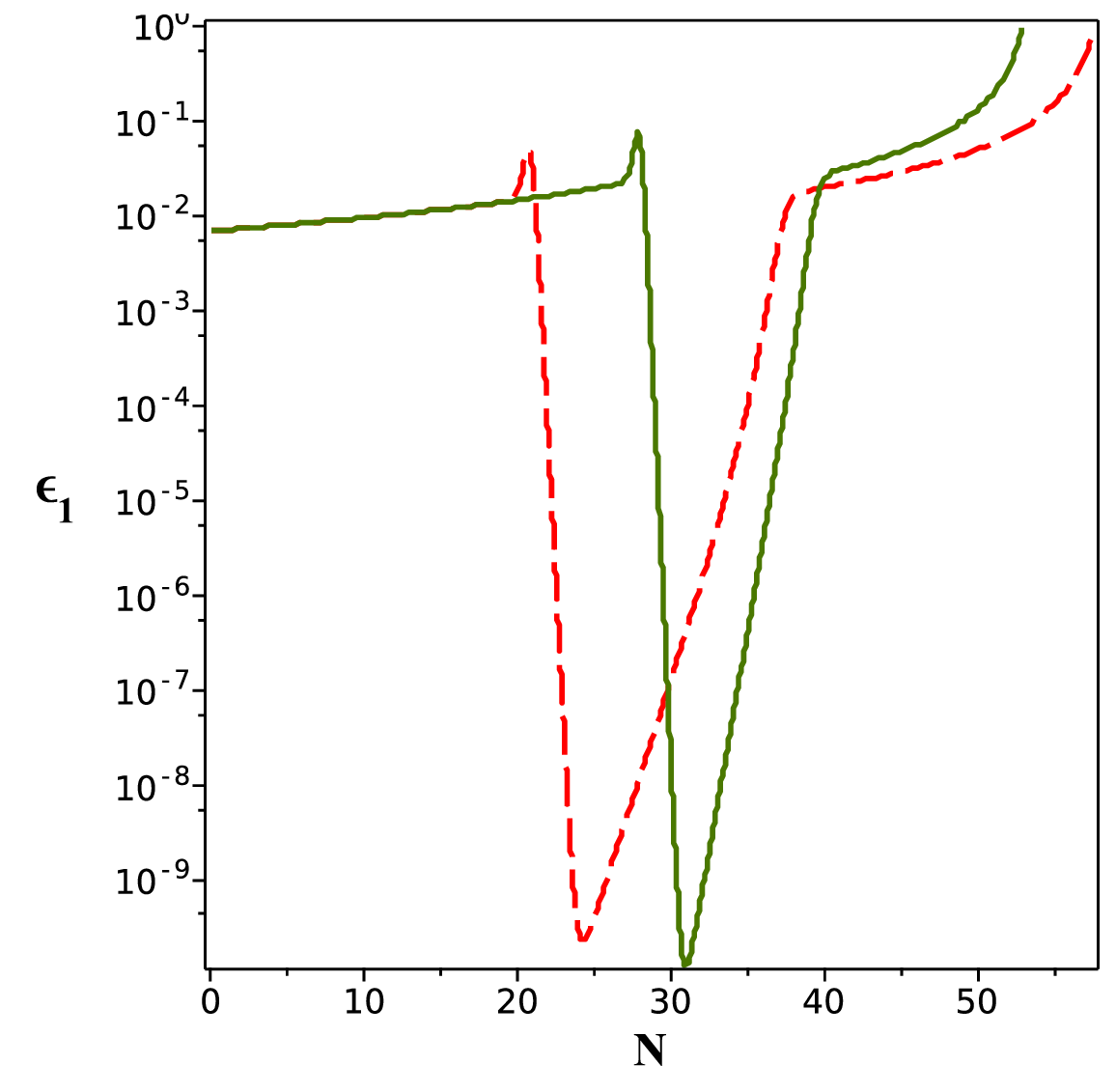}
		\caption{}
	\end{subfigure}
	\begin{subfigure}{0.40\linewidth}
		\centering
		\includegraphics[width=\linewidth]{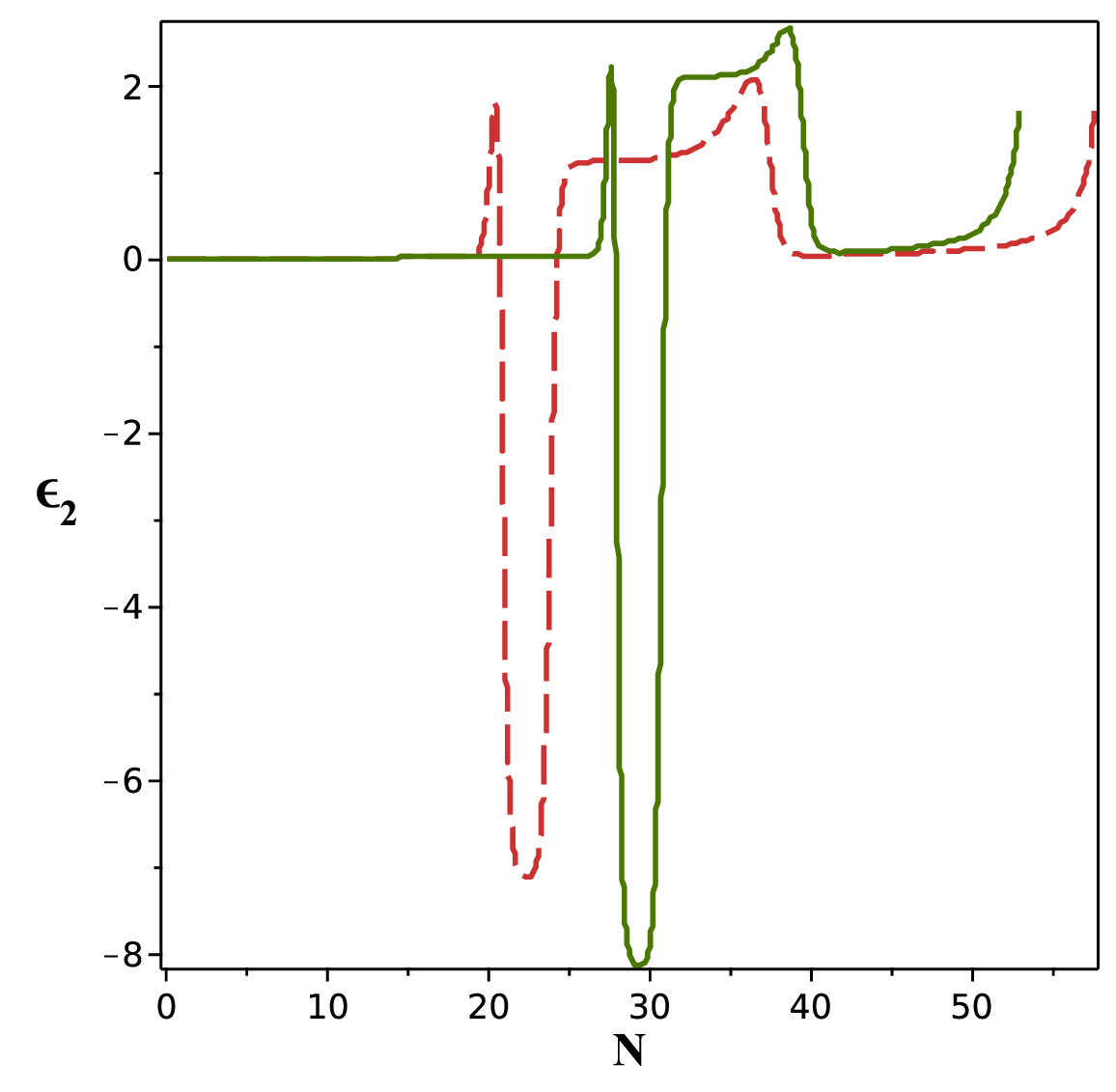}
		\caption{}
	\end{subfigure}
	\caption{Evolution of the slow-roll parameters for two representative realizations of the DBI model. The red dashed curves correspond to $A_{\rm V}=0.03115$ and $\phi_{0}/M_{\rm Pl}=8.409$, while the green solid curves correspond to $A_{\rm V}=0.04817$ and $\phi_{0}/M_{\rm Pl}=6.680$. The localized feature in the potential induces a sharp suppression in $\epsilon_1(N)$ and a large excursion in $\epsilon_2(N)$, signaling a transient departure from the slow-roll regime.	}
	\label{fig1}
\end{figure}

Figure~\ref{fig1} shows the evolution of the first two slow-roll parameters for two representative realizations of the DBI model. The red dashed curve corresponds to $
A_{\rm V}=0.03115$ and $\phi_{0}/M_{\rm Pl}=8.409,
$
while the green solid curve corresponds to
$
A_{\rm V}=0.04817$ and $\phi_{0}/M_{\rm Pl}=6.680.
$
All other parameters are kept fixed as listed in Table~\ref{tab1}. In both cases, the localized feature in the potential, together with the localized enhancement in the warp factor, dynamically produces a suppression of $\epsilon_1(N)$. This transient suppression signals a short departure from the standard slow-roll regime and is accompanied by a large excursion in $\epsilon_2(N)$ around the feature region. The location and depth of the dip are controlled mainly by the position and amplitude of the localized feature in the potential. As a result, changing $\phi_{0}/M_{\rm Pl}$ shifts the feature in e-fold time, while changing $A_{\rm V}$ modifies the strength of the suppression. This behavior is important for the generation of small-scale curvature perturbations. During the transient feature, the effective term $z''/z$ in the Mukhanov--Sasaki equation is modified, allowing selected modes to undergo amplification before freezing on super-horizon scales. Therefore, the enhancement of $\mathcal{P}_{\zeta}(k)$ arises dynamically from the specified DBI functions $V(\phi)$ and $f(\phi)$. The large excursion of $\epsilon_2(N)$ around the feature reflects the rapid variation of $\epsilon_1(N)$ and signals a short departure from the standard slow-roll regime.

\begin{table}[t]
	\centering
	\caption{Benchmark parameter values used in the numerical analysis of the DBI model with potential and warp-factor features.}
	\begin{tabular}{c c c}
		\hline\hline
		Parameter & Case A & Case B \\
		\hline
		$\beta M_{\rm Pl}$              & $0.060$ & $0.060$ \\
		$A_{\rm V}$       & $0.03115$ & $0.04817$ \\
		$\phi_{0}/M_{\rm Pl}$          & $8.409$ & $6.680$ \\
		$\sigma/M_{\rm Pl}$       & $0.10$ & $0.10$ \\
		$f_0 M_{\rm Pl}^{4}$               & $0.01$ & $0.01$ \\
		$A_f$                & $2000$ & $2000$ \\
		\hline\hline
	\end{tabular}
	\label{tab1}
\end{table}

\subsection{Primordial curvature power spectrum}
Having established the background evolution, we now turn to the computation of the primordial curvature power spectrum. The transient non-attractor phase identified in Fig.~\ref{fig1} is expected to leave a distinctive imprint on small scales, which we quantify by solving the Mukhanov--Sasaki equation for individual Fourier modes.

\begin{figure}[t]
	\centering
	\includegraphics[width=0.5\linewidth]{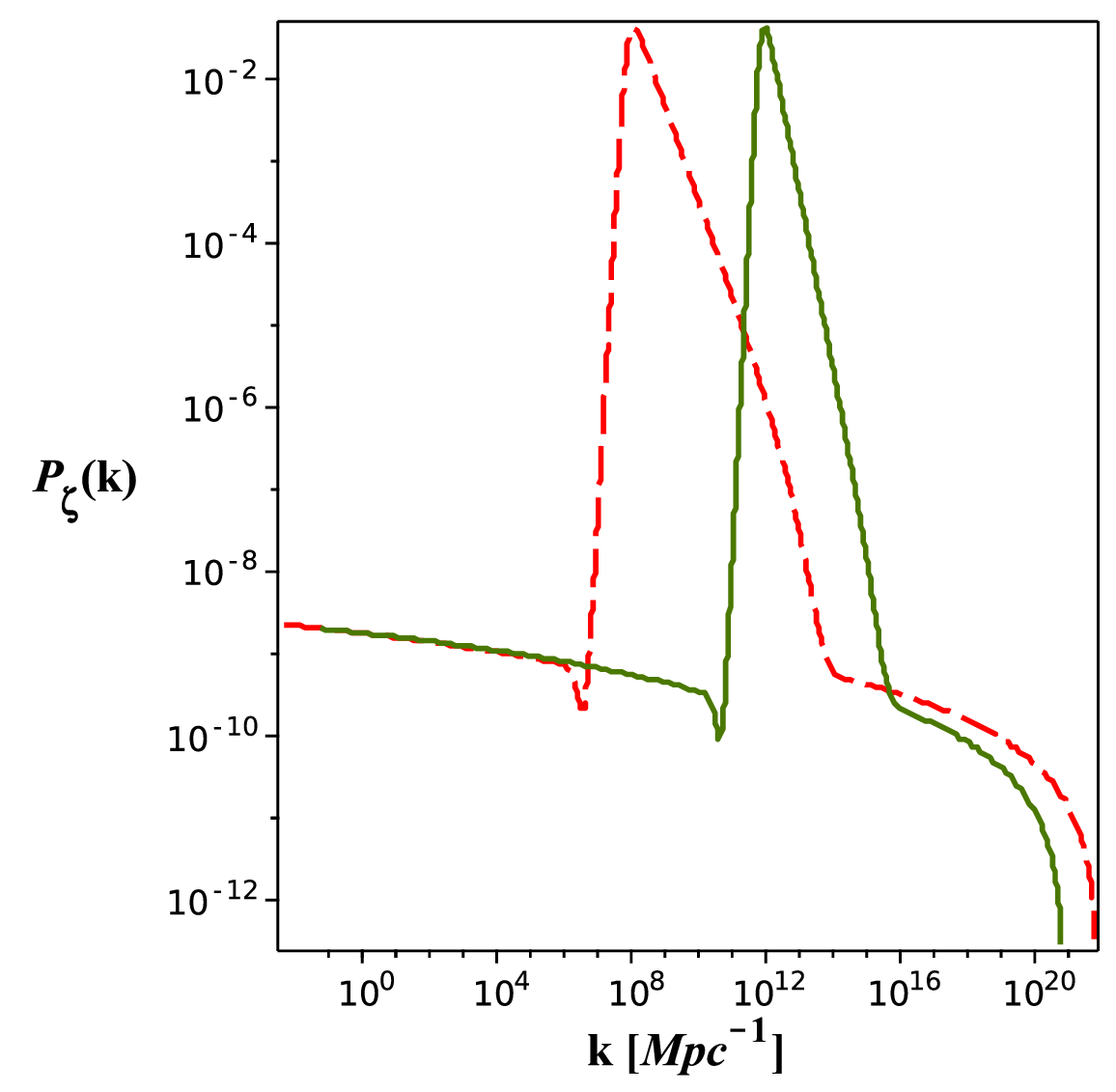}
	\caption{
Primordial curvature power spectrum $\mathcal{P}_{\zeta}(k)$ obtained by numerically solving the Mukhanov--Sasaki equation for two representative realizations of the DBI model. The red dashed curve corresponds to $A_{\rm V}=0.03115$ and $\phi_{0}/M_{\rm Pl}=8.409$, while the green solid curve corresponds to $A_{\rm V}=0.04817$ and $\phi_{0}/M_{\rm Pl}=6.680$. On large scales, the spectrum remains nearly scale invariant, whereas modes exiting the sound horizon during the transient feature exhibit a strong amplification, leading to a pronounced peak on small scales. The position and amplitude of the peak are controlled by the location and depth of the localized feature in the inflaton potential $V(\phi)$ and the warp factor $f(\phi)$.	}
	\label{fig2}
\end{figure}

We compute the primordial curvature power spectrum by solving the Mukhanov--Sasaki equation~\eqref{eq18} for each Fourier mode $k$. The evolution is initialized deep inside the sound horizon using the Bunch--Davies vacuum condition~\eqref{eq19}, and followed until the modes freeze on super-horizon scales. For each wavenumber, the power spectrum is evaluated at late times according to Eq.~\eqref{eq20}. The calculation is performed over a wide range of scales, extending from CMB-relevant modes to much smaller scales relevant for PBH formation. Figure~\ref{fig2} shows a representative example of the resulting primordial curvature power spectrum for two different choices of the potential feature parameters. On large scales, corresponding to modes exiting the horizon well before the feature, the spectrum remains nearly scale invariant. In contrast, modes exiting the sound horizon during the transient non-attractor phase experience significant amplification, leading to a pronounced peak on small scales. This enhancement originates from the localized structures in the inflaton potential $V(\phi)$ and the warp factor $f(\phi)$, which dynamically generate a transient suppression of $\epsilon_1(N)$ and a reduction of the sound speed. The position and amplitude of the peak are therefore controlled by the location, width, and depth of these features. A deeper and more prolonged suppression of $\epsilon_1$, together with a stronger reduction of the sound speed, leads to a more efficient amplification of curvature perturbations. The overall normalization of the spectrum is fixed by matching the amplitude at the pivot scale to the observed value $A_s$. This ensures that the enhancement on small scales arises from the underlying DBI dynamics rather than from an arbitrary rescaling. As a result, the power spectrum exhibits a localized and narrow peak, which plays a central role in determining the abundance and characteristic mass range of PBHs.

\subsection{PBH abundance}
\begin{figure}[t]
	\centering
	\includegraphics[width=0.65\linewidth]{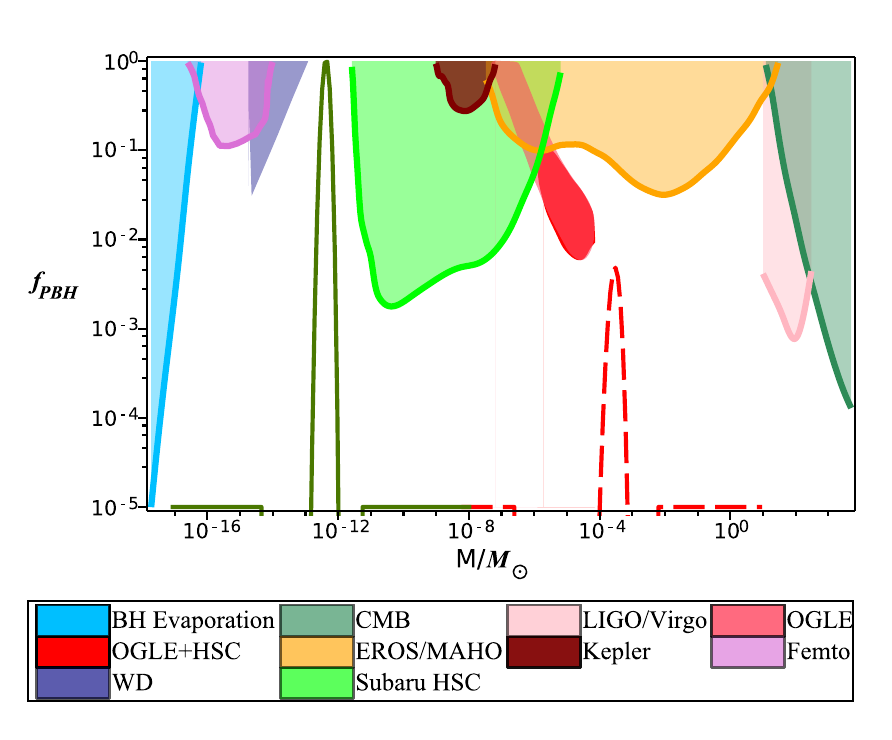}
\caption{
	PBH abundance $f_{\rm PBH}(M)$ as a function of mass for two representative realizations of the DBI model. The red dashed and green solid curves correspond to the two benchmark parameter choices considered in this work. The shaded regions indicate current observational constraints from evaporation, CMB, microlensing surveys (EROS/MACHO, Subaru HSC, OGLE), and gravitational wave observations (LIGO/Virgo and Kepler). The localized enhancement in the primordial power spectrum leads to a peaked PBH abundance, whose position in mass space is determined by the scale of the inflationary feature.}
\label{fig3}
\end{figure}

Figure~\ref{fig3} shows the resulting PBH abundance $f_{\rm PBH}(M)$ in the DBI scenario for two representative realizations (red dashed and green solid curves), together with current observational constraints. The shaded regions indicate excluded parameter space from various probes, including evaporation bounds~\cite{Car10}, CMB constraints~\cite{Ali17b,Pou17}, microlensing surveys (EROS/MACHO~\cite{Tis07}, Subaru HSC~\cite{Nii19}, OGLE~\cite{Nii19b}), and gravitational wave observations from LIGO/Virgo~\cite{Ali17} and Kepler~\cite{Gri13}. Owing to the localized enhancement of the primordial power spectrum, the PBH abundance exhibits a narrow and peaked structure. The two realizations correspond to peaks at different mass scales, reflecting the shift in the underlying inflationary dynamics. This behavior follows from the direct relation between the enhancement scale in $\mathcal{P}_{\zeta}(k)$ and the horizon mass at re-entry. As shown in Fig.~\ref{fig3}, the predicted PBH abundance is strongly scale-dependent and differs significantly between the two realizations. In one case (red dashed curve), the abundance reaches a peak value of order $f_{\rm PBH}\sim10^{-3}$-$10^{-2}$, corresponding to a subdominant PBH population. In contrast, the other realization (green solid curve) produces a much larger enhancement, with $f_{\rm PBH}$ approaching $\mathcal{O}(1)$ near the peak. This suggests that, for suitable parameter choices, PBHs could account for a significant fraction of dark matter within a narrow mass range. More generally, these results demonstrate that the DBI framework allows for a controlled localization of PBH production in mass space. By adjusting the position and strength of the inflationary feature, both the amplitude and the mass scale of the PBH abundance can be tuned, enabling a systematic exploration of different phenomenological regimes. 
We note that, although the PBH abundance is sensitive to the amplitude of the power spectrum, the localized enhancement mechanism considered here does not rely on a single finely tuned point in parameter space. Instead, a range of parameter values leads to qualitatively similar behavior, as also supported by the parameter scan discussed below.

\subsection{Induced gravitational wave spectrum}
\begin{figure}[t]
	\centering
	\includegraphics[width=0.65\linewidth]{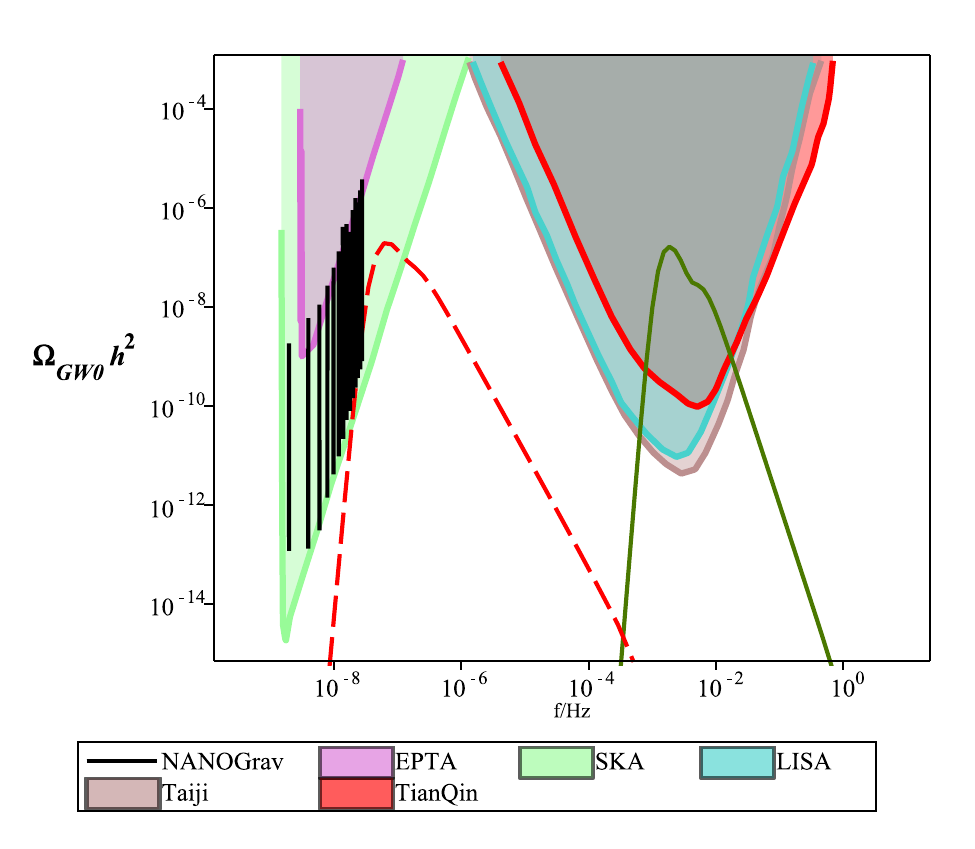}
	\caption{
	IGWs spectra generated by enhanced scalar perturbations in the DBI model. The red dashed and green solid curves correspond to the two benchmark realizations considered in this work. Shaded regions indicate the projected sensitivities of pulsar timing arrays (EPTA, SKA), space-based interferometers (LISA, Taiji, TianQin) and NANOGrav band.	}
		\label{fig4}
\end{figure}

The enhancement of primordial curvature perturbations generated by the localized features in the inflaton potential $V(\phi)$ and the warp factor $f(\phi)$ inevitably leads to the production of a stochastic background of IGWs at second order in cosmological perturbation theory. These tensor modes are sourced by enhanced scalar perturbations upon horizon re-entry during the radiation-dominated era. Using the numerically obtained primordial power spectrum $\mathcal{P}_{\zeta}(k)$, we compute the present-day energy density spectrum of IGWs, $\Omega_{\rm GW,0}(f)$, following the standard second-order formalism. The resulting spectra are shown in Fig.~\ref{fig4} for the two representative realizations considered in this work. In both cases, the IGWs spectrum exhibits a pronounced peak at frequencies corresponding to the horizon re-entry scale of the enhanced curvature perturbations. The position of the peak is determined by the characteristic scale of the enhancement in $\mathcal{P}_{\zeta}(k)$ and is therefore directly related to the corresponding PBH mass scale. We compare our predictions with the projected sensitivities of pulsar timing arrays (EPTA~\cite{Fer10,Hob10,McL13,Hob13}, SKA~\cite{Moo15}), space-based interferometers (LISA~\cite{Ama17,Dan97}, Taiji~\cite{Hu17}, TianQin~\cite{Luo16}) and NANOGrav 15-year data~\cite{NANO}. As shown in Fig.~\ref{fig4}, the two realizations peak in different frequency bands. The green solid curve lies in the frequency range relevant for future space-based interferometers, while the red dashed curve is shifted toward lower frequencies, closer to the pulsar-timing range. This illustrates that the observability of the IGWs signal is controlled not only by its amplitude, but also by the frequency location of the peak. We note that, for the specific benchmark parameter choices considered in this work, a small portion of the low-frequency tail of the red dashed spectrum partially overlaps with the NANOGrav 15-year data band. This overlap occurs only within a limited frequency interval and does not represent a full fit to the observed signal.

\subsection{CMB constraints in the \(r\)-\(n_s\) plane}
\begin{figure}[t]
	\centering
	\includegraphics[width=0.6\linewidth]{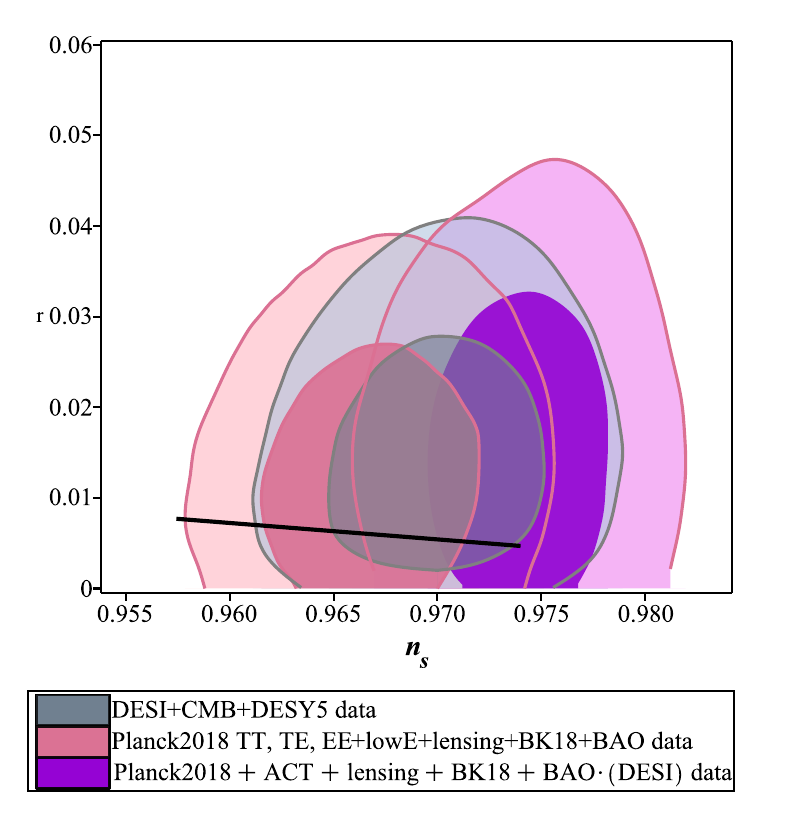}
	\caption{Predictions of the DBI model in the $r$–$n_s$ plane for the scanned parameter space. The black solid curve represents the model trajectory, while the shaded regions indicate observational constraints from Planck 2018, ACT, DES, BK18, and BAO data. }
	\label{fig5}
\end{figure}

In addition to producing an enhanced small-scale power spectrum and a viable PBH abundance, the model must also remain consistent with CMB constraints on the scalar spectral index and the tensor-to-scalar ratio. We therefore examine the predictions of the DBI model in the $r-n_s$ plane and compare them with current observational bounds. To this end, we perform a scan over the parameters of the potential feature, $0.0250 < A_{\rm V} < 0.0630$ and $6.00 < \phi_{0}/M_{\rm Pl} < 9.00$, while keeping the remaining parameters fixed to the representative values listed in Table~\ref{tab1}. These ranges are chosen to encompass the benchmark realizations considered in this work, ensuring that the scan probes the same region of parameter space that gives rise to the localized features discussed above. For each parameter set, the background equations are solved numerically and the overall normalization of the potential is fixed by matching the scalar amplitude at the pivot scale to $A_s\simeq 2.1\times10^{-9}$. The scalar spectral index and tensor-to-scalar ratio are then evaluated at the pivot scale using Eqs.~\eqref{eq13} and~\eqref{eq14}. The resulting predictions in the $r-n_s$ plane are displayed in Fig.~\ref{fig5}, together with observational constraints from Planck 2018, ACT, DES, BK18, and BAO data. The shaded contours correspond to the combined likelihood regions from Planck2018+ACT+lensing+BK18+BAO~\cite{ACT}, DESI+CMB+DESY5\cite{DESI}, and Planck2018 TT, TE, EE+lowE+lensing+BK18+BAO\cite{Pl18,pa22} analyses. 

As shown in Fig.~\ref{fig5}, the predictions lie within or close to the observationally allowed regions for a range of parameter values. The trajectory remains near the region preferred by current CMB and large-scale structure data. This indicates that the localized feature responsible for the small-scale enhancement of $\mathcal{P}_{\zeta}(k)$ can be implemented without significantly affecting the large-scale predictions of the model. The black solid curve in Fig.~\ref{fig5} illustrates the variation of these predictions under the scan of the feature parameters. Since the localized feature affects modes that exit the sound horizon well after the CMB pivot scale, its impact on large-scale observables is expected to be limited. As a result, the trajectory in the $r$–$n_s$ plane remains confined to the region favored by observations, while significant modifications can still occur on much smaller scales.	Consequently, the same DBI setup can generate an enhanced small-scale power spectrum, viable PBH abundances, and IGWs signals, while remaining compatible with current constraints in the $r$–$n_s$ plane within the explored parameter space.

\section{Conclusion}
\label{sec5}

In this work, we studied PBH formation in a DBI inflationary framework in which the background dynamics are determined by explicit functional forms of the inflaton potential $V(\phi)$ and the warp factor $f(\phi)$. By solving the background equations numerically, we obtained the evolution of the slow-roll parameters and the sound speed, which were then used to compute the primordial curvature perturbations. We showed that a localized feature in the potential, accompanied by a correlated structure in the warp factor, can dynamically induce a transient suppression of the slow-roll parameter. This leads to a short-lived non-attractor phase, during which curvature perturbations are efficiently amplified. As a result, the primordial power spectrum develops a narrow and pronounced peak on small scales, while remaining consistent with CMB-scale observations.

Using the enhanced power spectrum, we computed the resulting PBH abundance. The model produces localized peaks in the PBH abundance, whose position and amplitude are directly controlled by the scale and strength of the inflationary feature. This allows for a flexible realization of PBH production across different mass ranges within a single framework. We also evaluated the stochastic background of IGWs generated at second order by the enhanced scalar perturbations. The resulting spectra exhibit characteristic peaks whose frequencies are set by the scale of the enhancement in the curvature power spectrum. Depending on the model parameters, the predicted signals can lie in frequency bands relevant for pulsar timing arrays or future space-based interferometers. Finally, we examined the predictions of the model in the $r$–$n_s$ plane and found that they remain consistent with current observational constraints. This demonstrates that the mechanism responsible for small-scale amplification can operate without significantly affecting the large-scale predictions of the model.

Overall, the DBI framework considered here can provide a consistent and physically motivated setting in which localized features in the inflaton dynamics can simultaneously generate enhanced small-scale perturbations, PBHs, and IGWs signals, while remaining compatible with current cosmological observations.\\

\textbf{ACKNOWLEDGMENTS}\\
I thank the referee for the very insightful comments that have
improved the quality of the paper considerably.
\\

\textbf{Data Availability Statement:} All relevant analytical expressions and numerical results are included in the manuscript.

\textbf{Code/Software Availability Statement:}
	No public code repository is associated with this work.

\end{document}